\documentclass[pop,fleqn]{w-art}
\usepackage{times}
\usepackage{w-thm}
\usepackage{amsmath}
\usepackage{amsfonts} 
\usepackage{amssymb} 

\theoremstyle{plain}

\theoremstyle{definition}

\usepackage{graphicx}

\def\one{{\rm 1\kern -.9mm l}}

\begin{document}
%%    The information for the title page will be placed between
%%    \begin{document} and \maketitle. The order of most entries
%%    is determined by the class file and can not be changed by
%%    rearranging them. The maketitle command follows after the
%%    abstract.
%%
%%    Most of the following commands will be completed by the publisher.
%%
%%    The copyrightyear is defined in the .clo file as the first argument
%%    of the copyrightinfo command. If the copyrightyear differs from that
%%    value it might be adjusted by the following definition:
%%
%% \renewcommand{\copyrightyear}{2003}% uncomment to change the copyrightyear.
%%
\DOIsuffix{theDOIsuffix}
%%
%% issueinfo for header and copyright line
\Volume{51}
\Issue{1}
\Month{01}
\Year{2003}
%%
%%    First and last pagenumber of the article. If the option
%%    'autolastpage' is set (default) the second argument may be left empty.
\pagespan{1}{}
%%
%%    Dates will be filled in by the publisher. The 'reviseddate' and
%%    'dateposted' (Published online) entry may be left empty.
%\Receiveddate{15 November 2003}
%\Reviseddate{30 November 2003}
%\Accepteddate{1 December 2004}
%\Dateposted{3 December 2004}
%%
\keywords{Supermembrane, minimal immersions, M-theory, 4D compactifications}
\subjclass[pacs]{11.25.-w, 11.25.Yb, 11.25.Mj, 11.15.tk}

%% \pretitle{Editor's Choice}

%% We have a short and a long form for the title. The short form
%% (optional argument) goes into the running head.

\title [Minimally Immersed  4D Supermembrane]{The Minimally Immersed 4D  Supermembrane.}

%% Please do not enter footnotes or \inst{}-notes into the optional
%% argument of the author command. The optional argument will go into
%% the header.  If there is only one address the marker \inst{x} may be
%% omitted.

%% Information for the first author.
\author[M.P. Garcia del Moral]{M. P. Garc\'\i a del Moral\inst{1}%
  \footnote{ E-mail:~\textsf{garcia@to.infn.it}}}
\address[\inst{1}]{Dip. di Fisica Teorica, Universit\`a di Torino
and I.N.FN., sez. di Torino\\ 
Via Pietro Giuria 1, 10125 Torino (Italy)}
%%
%%    Information for the second author
\author[J.M. Pena]{J. M. Pena \inst{2} \footnote{E-mail:~\textsf{jpena@fisica.ciens.ucv.ve, arestu@usb,ve}}}
\address[\inst{2}]{ Departamento de F\'\i sica, Facultad de Ciencias,
 Universidad Central de Venezuela,\\
 A.P. 47270, Caracas 1041-A, (Venezuela)}
%\address[\inst{2}]{Second address}
%%
%%    Information for the third author
\author[A. Restuccia]{A. Restuccia\inst{3} \footnote{E-mail:~\textsf{restucci@aei.mpg.de}}}
\address[\inst{3}]{Max-Planck-Institut f\"ur Gravitationphysik, Albert-Einstein-Institut\\
M\"ulenberg 1, D-14476 Potsdam, Germany $\&$
Departamento de F\'\i sica, Universidad Sim\'on Bol\'\i var\\
Apartado 89000, Caracas 1080-A, (Venezuela)}
%%
%%    \dedicatory{This is a dedicatory.}
\begin{abstract}
 In this note we summarize some of the properties found in \cite{joselen}, and its relation with \cite{g2}. We comment on the construction of the action of the 11D supermembrane with nontrivial central charges
minimally immersed on a 7D toroidal manifold is obtained (MIM2).The transverse coordinates to the supermembrane are maps to a 4D Minkowski space-time. The action is
invariant under additional symmetries in comparison to the
supermembrane on a 11D Minkowski target space. The hamiltonian in the
LCG is invariant under conformal transformations on the Riemann
surface base manifold.  The spectrum of the regularized hamiltonian
is discrete with finite multiplicity. Its resolvent is compact. Susy
is spontaneously broken, due to the topological central charge
condition, to four supersymmetries in 4D, the vacuum belongs to an
N=1 supermultiplet. When assuming the target-space to be an
isotropic 7-tori, the potential does not contain any flat direction,
it is stable on the moduli space of parameters. Moreover due to the discrete symmetries of the hamiltonian, there are only 7 possible minimal holomorphic immersions of the MIM2 on the 7-torus. When these symmetries are identified on the target space, it corresponds to compactify the MIM2 on a orbifold with G2 structure. Once the singularities are resolved it leads to the compactification of the MIM2 on a G2 manifold as shown in \cite{g2}. 
\end{abstract}
%% maketitle must follow the abstract.
\maketitle                   % Produces the title.

%% If there is not enough space inside the running head
%% for all authors including the title you may provide
%% the leftmark in one of the following three forms:

%% \renewcommand{\leftmark}
%% {First Author: A Short Title}

%% \renewcommand{\leftmark}
%% {First Author and Second Author: A Short Title}

%% \renewcommand{\leftmark}
%% {First Author et al.: A Short Title}

%% \tableofcontents  % Produces the table of contents.

\section{Introduction}
There have been interesting advances towards the quantization
(\cite{torrealba}-\cite{bgmr2})of a sector of M-theory, the 11D minimally immersed supermembrane (MIM2), that may also provide tools to attack the more general problem of M-theory quantization. The goal to achieve is, departing from 11D, to obtain a consistent quantum theory 
in 4D with an $N=1$ or $N=0$ supersymmetries, moduli free, in
agreement with the observed 4D physics. Attempts to formulate the
theory in 4D have been done mainly in the supergravity approach
\cite{gutowski, lukas}, including fluxes \cite{acharya, dall'agata},
but so far no exact formulation has been found. An effective theory when
compactified to a 4D model contains many vacua due to the presence
of moduli fields. Stabilizacion of these moduli  is an important
issue to be achieved. For some interesting proposals from M-theory see
 \cite{acharya, diana}. In the following we will consider the $11D$
M2-brane with irreducible wrapping on the compact sector of the
target manifold \cite{joselen}-\cite{bgmr2} ( MIM2). This implies the existence of a non
trivial central charge in the supersymmetric algebra. Its spectral
properties have been obtained in several papers
\cite{gmr}-\cite{bgmr2}. The MIM2 contains the information about bound states of Dbranes, for example when compactified on 9D, it is  equivalent to a bundle of D2-D0 branes. The MIM2 also contains inside its spectrum nonperturbative  string states like (F,Dp) branes and it has been proved in  \cite{gmmr} to be the 11D origin of the SL(2,Z) multiplets of IIB theory and it may also be the  11D origin of the nonperturbative dyonic states of type IIA \cite{gmmr}that cannot be seen at perturbative level. This fact may be of relevance since Dp-branes are interesting since they allow to obtain nonabelian gauge groups and are able to reproduce semirealistic models of phenomenology and cosmology. Recently  bound states of Dbranes have also enter into the game since they are able to capture nonperturbative effects that could explain very tiny effects inside MSSM in a natural way, i.e. smallness of neutrino masses or Yukawa couplings.
The purpose of this note is to show the main aspect of the construction of  the action for the
supermembrane with nontrivial central charges compactified on a
$T^{7}$ realized through all of the allowed holomorphic minimal immersions and analyze its physical properties.
\section{D=11 Supermembrane with  central charges on a $M_{5}\times T^{6}$ target manifold}
The hamiltonian of the $D=11$ Supermembrane \cite{bst} may be
defined in terms of maps $X^{M}$, $M=0,\dots, 10$, from a base
manifold $R\times \Sigma$, where $\Sigma$ is a Riemann surface of
genus $g$ onto a target manifold which we will assume to be $11-l$
Minkowski $\times$ $l$-dim Torus. The canonical reduced hamiltonian
to the light-cone gauge has the expression

\begin{equation}\label{e1}
   \int_\Sigma  \sqrt{W} \left(\frac{1}{2}
\left(\frac{P_M}{\sqrt{W}}\right)^2 +\frac{1}{4} \{X^M,X^N\}^2+
{\small\mathrm{\ Fermionic\ terms\ }}\right)
\end{equation}
subject to the constraints
\begin{equation}  \label{e2}
\phi_{1}:=d(\frac{p_{M}}{\sqrt{W}}dX^{M})=0
\end{equation}
and
\begin{equation} \label{e3}
\phi_{2}:=
   \oint_{C_{s}}\frac{P_M}{\sqrt{W}} \quad dX^M = 0,
\end{equation}
where the range of $M$ is now $M=1,\dots,9$ corresponding to the
transverse coordinates in the light-cone gauge, $C_{s}$,
$s=1,\dots,2g$ is a basis of  1-dimensional homology on $\Sigma$,
 \begin{equation} \label{e4}\{X^{M}, X^{N}\}= \frac{\epsilon
^{ab}}{\sqrt{W(\sigma)}}\partial_{a}X^{M}\partial_{b}X^{N}. \end{equation}
$a,b=1,2$ and $\sigma^{a}$ are local coordinates over $\Sigma$.
$W(\sigma)$ is a scalar density introduced in the light-cone gauge
fixing procedure. $\phi_{1}$ and $\phi_{2}$ are generators of area
preserving diffeomorphisms. That is \begin{equation}
\sigma\to\sigma^{'}\quad\to\quad W^{'}(\sigma)= W(\sigma).\nonumber
\end{equation}
The $SU(N)$ regularized model obtained from (\ref{e1}) \cite{dwhn} was shown to
have conti\-nuous spectrum from $[0,\infty)$,
\cite{dwln},\cite{dwmn},\cite{dwhn}.
In what follows we will impose a topological restriction on the
configuration space. It characterizes a $D=11$ supermembrane with
non-trivial central charges generated by the wrapping on the compact
sector of the target space
\cite{gmr},\cite{bgmmr},\cite{bgmr},\cite{bgmr2}. Following
\cite{bellorin} we may extend the original construction on a
$M_{9}\times T^{2}$ to $M_{7}\times T^{4}$, $M_{5}\times T^{6}$
target manifolds by considering genus $1,2,3$ Riemann surfaces on
the base respectively. We are interested in reducing the theory to a
4 dimensional model, we will then assume a target manifold
$M_{4}\times T^{6}\times S^{1}$. The configuration maps satisfy:
\begin{equation}\label{e5}
 \oint_{c_{s}}dX^{r}=2\pi S_{s}^{r}R^{r}\quad r,s=1,\dots,6, \quad
 \oint_{c_{s}}dX^{m}=0 \quad m=8,9
 \end{equation}
and
\begin{equation}
\label{e6}
\oint_{c_{s}}dX^{7}=2\pi L_{s}R, 
\end{equation}
where $S^{r}_{s}, L_{s}$ are integers and $R^{r}, r=1,\dots,6$ are
the radius of $T^{6}=S^{1}\times\dots\times S^{1}$ while $R$ is the
radius of the remaining $S^{1}$ on the target.
We now impose the central charge condition \begin{equation}\label{e8}
I^{rs}\equiv \int_{\Sigma}dX^{r}\wedge dX^{s}=(2\pi
R^{r}R^{s})\omega^{rs}n \end{equation} where $\omega^{rs}$ is a symplectic
matrix on the $T^{6}$ sector of the target and $n$ denotes an integer representing the irreducible winding. The topological
condition (\ref{e8}) does not change the field equations of the
hamiltonian (\ref{e1}). In addition to the field equations obtained
from (\ref{e1}), the classical configurations must satisfy the
condition (\ref{e8}). In the quantum theory, the space of physical
configurations is also restricted by the condition
(\ref{e8}) \cite{torrealba},\cite{ovalle}.\\

We consider now the most general map satisfying condition
(\ref{e8}): \begin{equation} dX^{r}=M_{s}^{r}d\widehat{X}^{s}+dA^{r} \end{equation} where
$d\widehat{X}^{s}$, $s=1,\dots,2g$ is a basis of harmonic one-forms
over $\Sigma$ and impose the constraints (\ref{e2}),(\ref{e3}). It turns out that $M_{s}^{r}$ can be expressed in terms of a matrix $S\in
Sp(2g,Z)$,\cite{joselen}.
The natural election for $\sqrt{W(\sigma)}$ in this geometrical
setting is define \begin{equation}
\sqrt{W(\sigma)}=\frac{1}{2}\partial_{a}\widehat{X}^{r}\partial_{b}\widehat{X}^{s}\omega_{rs}.
\end{equation}
$\sqrt{W(\sigma)}$ is then invariant under the change \begin{equation}
d\widehat{X}^{r}\to S_{s}^{r}d\widehat{X}^{s}, \quad S\in Sp(2g,Z)
\end{equation}
We thus conclude that the theory  is invariant not only under the
diffeomorphisms generated by $\phi_{1}$ and $\phi_{2}$ but also
under the diffeomorphisms, biholomorphic maps, changing the
canonical basis of homology by a modular transformation. The theory
of supermembranes with central charges in the light cone gauge (LCG)
we have constructed depends then on the moduli space of
compact Riemanian surfaces $M_{g}$ only. In addition when compactify in 9D there has been proved in \cite{gmmr} the hamiltonian is also invariant under  a second SL(2,Z) symmetry associated to the $T^{2}$ target space that transform the Teichmuller parameter of the 2-torus.$T^{2}$.
\section{Compactification on the remaining $S^{1}$}
We will discuss two approaches for the analysis of the
compactification on the remaining $S^{1}$.
In the first case, we may solve the condition (\ref{e6}), we obtain
 \begin{equation}\label{e10}
 dX^{7}= R L_{s}d\widehat{X}^{s}+d\widehat{\phi}
 \end{equation}
 where $d\widehat{\phi}$ is an exact 1-form and $d\widehat{X}^{s}$
 as before are a basis of harmonic 1-forms over $\Sigma$.
We may analyze the contribution of the $dX^{7}$ field to the potential of the hamiltonian, we call it $V_{7}$. It is bounded from below \begin{equation}\label{e15}
 V_{7}\geq \left\langle
(\mathcal{D}_{r}\phi)^{2}+\{X^{m},X^{7}\}^{2}\right\rangle \end{equation}
which directly shows that the winding corresponding to $dX^{7}$ does
not affect the qualitative properties of the spectrum of the hamiltonian \cite{bellorin}.
The inequality (\ref{e15}) will ensure that the discretness
property of the latest is also valid for the original complete
hamiltonian.
We will assume the dual formulation  to the hamiltonian \cite{joselen}
 when $dX^{7}$ is restricted by the condition (\ref{e6})
ensuring that $X^{7}$ takes values on $S^{1}$. We follow
\cite{caicedo}.
We notice that $A_{s}$ is not a connection in a line bundle over
$\Sigma$. In fact the condition
\begin{equation}
\int_{\Sigma}F_{ab}d\sigma^{a}\wedge d\sigma^{b}=2\pi n
\end{equation}
is not necessarily satisfied. In order to have a connection on line
bundle over $\Sigma$ one should require a periodic euclidean time on
the functional integral formulation. In that case the condition
(\ref{e6}), ensures that $F_{\mu\nu}$ is the curvature of a one-form
connection over the three dimensional base manifold. Under this
assumption the condition (\ref{e6}) for any $L_{s}$ implies
summation over all $U(1)$ principle bundles.

The final expression of the dual formulation of the hamiltonian when
$X^{7}$ is wrapped on a $S^{1}$, condition (\ref{e6}), is
\begin{equation}\label{e}
\begin{aligned}
H_{d}=&\int \sqrt{w}d\sigma^{1}\wedge d\sigma^{2}[\frac{1}{2}(\frac{P_{m}}{\sqrt{W}})^{2}
+\frac{1}{2}(\frac{\Pi^{r}}{\sqrt{W}})^{2}+\frac{1}{4}\{X^{m},X^{n}\}^{2}+\frac{1}{2}(\mathcal{D}_{r}X^{m})^{2}\\
& \nonumber +\frac{1}{4}(\mathcal{F}_{rs})^{2}+\frac{1}{2}(F_{ab}\frac{\epsilon^{ab}}{\sqrt{W}})^{2}
+\frac{1}{8}(\frac{\Pi^{c}}{\sqrt{W}}\partial_{c}X^{m})^{2}+\frac{1}{8}[\Pi^{c}\partial_{c}(\widehat{X}_{r}+A_{r})]^{2}]+\\
& \nonumber \Lambda(\{\frac{P_{m}}{\sqrt{W}},X^{m}\}-\mathcal{D}_{r}\Pi^{r}
-\frac{1}{2}\Pi^{c}\partial_{c}(F_{ab}\frac{\epsilon^{ab}}{\sqrt{W}}))+\lambda\partial_{c}\Pi^{c}]+ susy
\end{aligned}
\end{equation}
where  $\mathcal{D}_r X^{m}=D_{r}X^{m} +\{A_{r},X^{m}\}$,
$\mathcal{F}_{rs}=D_{r}A_s-D_{s }A_r+ \{A_r,A_s\}$, \\
 $D_{r}=2\pi
R^{r}\frac{\epsilon^{ab}}{\sqrt{W}}\partial_{a}\widehat{X}^{r}\partial_{b}$
and $P_{m}$ and $\Pi_{r}$ are the conjugate momenta to $X^{m}$ and
$A_{r}$ respectively. $\mathcal{D}_{r}$ and $\mathcal{F}_{rs}$ are
the covariant derivative and curvature of a symplectic
noncommutative theory \cite{ovalle},\cite{bgmmr}, constructed from
the symplectic structure $\frac{\epsilon^{ab}}{\sqrt{W}}$ introduced
by the central charge. The integral of the curvature we take it to be
constant and the volume term corresponds to the value of the
hamiltonian at its ground state. The physical
degrees of the theory are the $X^{m}, A_{r},X_{7}$  together with its supersymmetric extension. They are
single valued fields on $\Sigma$.
\section{$N=1$ supersymmetry}
The topological condition associated to the central charge
determines an holomorphic minimal immersion from the $g$-Riemann
surface to the 2g-torus target manifold. This minimal immersion is
directly related to the BPS state that minimizes the hamiltonian.
When we start with the $g=1$ and $T^{2}$ on the target space the
ground state preserves $\frac{1}{2}$ of the original supersymmetry
with parameter a 32-component Majorana spinor. When we consider our
construction for a $g=2,3$ and $T^{4}, T^{6}$ torus on the target,
the analysis of the SUSY preservation becomes exactly the same as
when considering orthogonal intersection of 2-branes with the time
direction as the intersecting direction \cite{smith}. The SUSY of
the ground state preserves $\frac{1}{4}$, $\frac{1}{8}$ of the
original SUSY. The preservation of the ground state implies the
breaking of the supersymmetry. In the light cone gauge, we end up,
when $g=3$, with $\frac{1}{8}$ of the original SUSY, that is one
complex grassmann parameter corresponding to a $N=1$ light-cone SUSY
multiplet.
The action is invariant under the whole light-cone SUSY. However
when the vacuum is spontaneously fixed to one of them, the SUSY is
broken at the quantum level up to $N=1$ when the target is
$M_{5}\times T^{6}$. There is no further breaking when we compactify
the additional $S^{1}$, to have a target $M_{4}\times T^{6}\times
 S^{1}$.
\section {Discretness of the spectrum}
We consider a gauge fixing procedure on a BFV formulation of the
theory. We consider, in the usual way, a decomposition of all scalar
fields over $\Sigma$ in terms of an orthonormal discret basis
$Y_{A}(\sigma^{1},\sigma^{2})$ and
\begin{equation*}
\int_{\Sigma}\{Y_{A},Y_{B}\}Y_{C}=f_{ABC},
\end{equation*}
$f_{ABC}$ is consequently completely antisymmetric. We then replace
those expressions into the hamiltonian density and integrate the
$\sigma^{1},\sigma^{2}$ dependence. We obtain then a formulation of
the operator in terms of the $\tau$ dependent modes only. We now
consider a truncation of the operator, that is we restrict the range
of the indices $A,B,C$ to a finite set $N$ and introduce constants
$f_{AB}^{N\quad C}$ such that
\begin{equation}
lim_{N\to\infty}f_{AB}^{N\quad C}=f_{AB}^{C}
\end{equation}
$f_{AB}^{N\quad C}$ are the structure constants of $SU(N)$.
 In \cite{bgmr}
the  truncated supermembrane with central charges compactified on a $T^{2}$ was shown to have a  $SU(N)$ a
gauge symmetry. The algebra of first class constraints is isomorphic to the algebra of a $SU(N)$ gauge theory. We proceed to the analysis of the spectrum of the
truncated Schr\"oedinger operator associated to $\widehat{H}$
without further
requirements on the constants $f^{N}$:\\

i) The potential of the Schr\"oedinger  operator only vanishes at
the origin of the configuration space, \\
ii) There exists a constant
$M> 0$ such that \begin{equation} V(X,A,\phi)\ge M\vert\vert
(X,A,\phi))\vert\vert^{2} \end{equation}

The Schr\"oedinger operator is then bounded by an harmonic
oscillator. Consequently it has a compact resolvent. We now use
theorem 2 \cite{br} to show that: i) The ghost and antighost
contributions to the effective action assuming a gauge fixing condition linear on the configuration variables, ii) the fermionic
contribution to the susy hamiltonian, do not change the qualitative
properties of the spectrum of the hamiltonian. The regularized hamiltonian compactified on the target
space $M_{4}\times T^{6}\times S^{1}$ has then a compact resolvent and
hence a discrete spectrum with finite multiplicity. We expect the
same result to be valid for the exact theory.
\section{Physical properties}
Another of the characteristics of the theory is that due to the
topological condition the fields acquire mass. In here, the fields
of the theory $X^{m},A_{r},\phi$ acquire mass via the vector fields
$\widehat{X}_{r}$ defined on the supermembrane. There is no
violation of Lorentz invariance. It is important to point out that
the number of degrees of freedom in 11D and in 4D is preserved, but
just redistributed. This fact has the a advantage for many
phenomenological purposes of mantaining the number of fields
small.
At classical level, generically the analysis of moduli fields have
been performed in a supergravity approach
\cite{acharya}-\cite{diana}. Since our approach is exact these terms
do not appear, however the action posseses  scalars that may lead to
flat directions in the potential. We are going to analyze the two
types of classical moduli. This decoupling approach is only
justified iff the scales of stabilization (the masses of the moduli)
are clearly different, otherwise the minimization with respect to
the whole set of moduli (geometrical and of matter origin) should be
performed. The theory does not contain any string-like
configuration, this is because the scalar fields parametrizing the position of the supermmbrane gets all mass, so these type of moduli gets fixed. With respect to the geometrical moduli parametrizing the manifold if we assume that we compactify on a 7 isotropic tori it can be rigurosuly proved that  all of the moduli gets fixed.
An heuristical argument to understand better this effect of moduli stabilization is the
following: We are dealing in our construction with nontrivial gauge
bundles that can be represented as worldvolume fluxes
\cite{d2-d0}. Since for construction the mapping represent a
minimal immersion on the target space they induce a similar effect
that the one induced by the generalized calibration, that is, there is an associated flux effect on the target space. Minimal
immersions take also into account the dependence on the base
manifold, the Rieman surface chosen $\Sigma$. The condition of the
generalized calibration -which shows the deformation of the cycles
that are wrapped by the supermembrane- represent a condition for
minimizing the energy \cite{evslin}. It happens the same with the
minimal inmersions. For a given induced flux, one may expect the
volume to be fixed \cite{gm}.
\section{Conclusion}
We obtained the action of the D=11 supermembrane compactified on $T^6\times S^1$ 
with nontrivial central charge induced by a topological condition  invariant under  
supersymmetric and kappa symmetry transformations. The  hamiltonian in the LCG is
 invariant under conformal  transformations on the Riemann surface base manifold. 
The susy is spontaneously broken, by the vacuum
to  $1/8$ of the original one. It corresponds in 4D to a $N=1$ multiplet. Classicaly 
the hamiltonian does not contain singular configurations and at the quantum level the 
regularized hamiltonian has a discrete spectrum, with finite multiplicity. Its
 resolvent is compact. The potential does not contain any flat direction on 
configuration space nor on the moduli space of parameters. The hamiltonian is 
stable on both spaces. It is stable as a Schrodinger operator on configuration 
space and it is structurable stable  on the moduli space of parameters.
The discrete symmetries of the theory restrict the allowed minimal immersions to those corresponding to an orbifold with G2 structure. When the symmetries are identified on the target space they lead to a compactification on a true G2 manifold  \cite{g2}.
\section{Acknowledgements}
MPGM thanks the organizers of the Workshop ``ForcesUniverse 2007'' celebrated in Valencia, Spain for the possibility to present this work . We also would like to M.~Billo for his help.  M.P.G.M. is
partially supported by the European Comunity's Human Potential
Programme under contract MRTN-CT-2004-005104 and by the Italian MUR
under contracts PRIN-2005023102 and PRIN-2005024045. J.M.~Pena
research is carried by the Ph.D grant 'programa Mision Ciencia',
Caracas, Venezuela. The work of A. R. is partially supported by a grant from MPG, Albert Einstein Institute, Germany and  by PROSUL, under contrat CNPq 490134/2006-08.

\end{document}